\documentclass[a4paper,11pt]{article}
\usepackage{jheppub} 
 \usepackage{amssymb,amsthm}
 \usepackage{amsmath,amssymb,amsfonts,bm,amscd}
 \usepackage{graphicx}
 
 \usepackage{caption, subcaption}
 
 \usepackage{tikz}
 
\def\beq#1\eeq{\begin{align}#1\end{align}}

\usepackage{enumitem}

\title{Soft supersymmetry breaking of 4d $\mathcal{N}=2$ SCFT}

\author[a,b]{Dan Xie}

\affiliation[a]{Yau Mathematics Science center, Tsinghua University, Beijing, 10084, China}
\affiliation[b]{Department of Mathematics, Tsinghua University, Beijing, 10084, China}

\abstract{A classification of soft SUSY breaking deformation of general  four dimensional $\mathcal{N}=2$ SCFT is provided. Given the 
large class of newly discovered $\mathcal{N}=2$ SCFTs and their known properties such as the central charges and full information of BPS operators,  it is possible to get a huge number of new $\mathcal{N}=1$ SCFTs 
and non-supersymmetric CFTs. Many properties of these new $\mathcal{N}=1$ SCFTs such as central charges, chiral spectrum and Seiberg duality  can be  derived from known information of parent  $\mathcal{N}=2$ SCFT. }

\begin{document} 
\maketitle
\flushbottom

\section{Introduction}
Given a conformal field theory, it is crucial to understand its behavior under various deformations such as relevant deformations, exact marginal deformations, and
deformations by turning on expectation value of some operators. For a superconformal field theory (SCFT), one can completely classify supersymmetry (SUSY) preserving relevant or exact marginal 
deformation by using the representation theory of superconformal algebra \cite{Cordova:2016xhm}. The SUSY preserving deformation derived from turning on expectation value of operators is more complicated and one need to determine
the algebra structure involving BPS operators \footnote{A typical example is chiral ring structure of four dimensional $\mathcal{N}=1$ SCFT \cite{Cachazo:2002ry}.}.
The operator contents of a specific model and the infrared behavior after deformations, on the other hand, are much more difficult to study and  the solution often involves the understanding of
many highly nontrivial dynamical aspects of a theory. 

It is now clear that the space of four dimensional  $\mathcal{N}=2$ SCFTs is extremely large \cite{Gaiotto:2009we,Xie:2012hs,Wang:2015mra,Wang:2018gvb,Xie:2015rpa}.  These SCFTs are mostly strongly coupled and do not admit a conventional Lagrangian description. One can, however, learn many 
highly nontrivial properties about these theories by using powerful string theory and geometric methods. In particular, a lot of  knowledge about the BPS multiplets is known so we do know the existence of many interesting operators.
The operator contents of strongly coupled theory are often richer and more generic than the SCFT defined using Lagrangian description. For example, the lower bound 
of Coulomb brach operator is two  for a $\mathcal{N}=2$ SCFT which admits a  Lagrangian description; but for Argyres-Douglas type theory \cite{Argyres:1995jj,Argyres:1995xn}, the lower bound  can be actually arbitrarily close to one (which is the unitarity bound). So the deformation theory of these strongly coupled SCFTs is much richer. 
The behavior of $\mathcal{N}=2$ preserving deformations is well studied: the solution of the Coulomb branch is captured by finding a Seiberg-Witten geometry; and the Higgs branch solution or more generally the Schur sector can be solved by finding an associated 2d vertex operator algebra \cite{Beem:2013sza,Xie:2019zlb}.

The $\mathcal{N}=1$ preserving deformation of a $\mathcal{N}=2$ SCFT is also studied, and the main focus is on using Coulomb branch operators with scaling dimension two \cite{Benini:2009mz,Gadde:2013fma,Maruyoshi:2013hja} and the Higgs branch operators which also has scaling dimension two \cite{Gadde:2013fma,Maruyoshi:2013hja,Maruyoshi:2016aim,Agarwal:2016pjo}. 
$\mathcal{N}=1$ preserving deformation for simplest Argyres-Douglas theories are studied in \cite{Bolognesi:2015wta,Xie:2016hny,Buican:2016hnq}, where Coulomb branch operators with fractional scaling dimension are used. The SUSY breaking deformation is rarely studied though. 

The main purpose of this note is to initiate a systematical study of $\mathcal{N}=1$ preserving and SUSY breaking deformation of general four dimensional $\mathcal{N}=2$ 
SCFT. Instead of studying arbitrary SUSY breaking deformations of a SCFT, we focus on a special class of deformation called soft SUSY breaking deformation. Such deformation has been studied in the context of SUSY breaking 
model building, and has many interesting features \cite{Luty:2005sn}. In our context, the soft SUSY breaking deformations are defined as follows: one start with a SUSY preserving relevant or marginal deformation, and then promote 
the coupling constant to appropriate supermultiplet. 

We start with a classification of soft SUSY breaking of four dimensional $\mathcal{N}=1$ SCFT: a): F term deformation using chiral multiplet ${\cal B}_{r,(j_1,0)}$ with $r\leq 2$; b): D term deformation using conserved current multiplet $\hat{{\cal C}}_{(0,0)}$. For $\mathcal{N}=2$ SCFT, one can have F term soft SUSY breaking deformation by using: a): chiral multiplet ${\cal E}_{r,(0,0)}$ with $r\leq 2$; b):
$\hat{{\cal B}}_1$ multiplet. If we regard $\mathcal{N}=2$ SCFT as a $\mathcal{N}=1$ SCFT, we have more choices of deformations: firstly the constraint on $r$ charge on $\mathcal{N}=2$ chiral multiplet ${\cal E}_{r,(0,0)}$ is relaxed to $r<3$; secondly, 
$\hat{{\cal B}}_1$ multiplet contains a $\mathcal{N}=1$ conserved current multiplet $\hat{{\cal C}}_{(0,0)}$, and one can use it to do the deformation; thirdly, $\mathcal{N}=2$ supercurrent multiplet contains a $\mathcal{N}=1$ conserved current $\hat{{\cal C}}_{(0,0)}$ and one can use it to get a soft SUSY breaking deformation. 
Some of most interesting deformations are summarized in table. {\ref{deform}.

We do not attempt to study the IR behavior of  general soft deformations in this paper, here we only point one interesting feature of these deformations. 
Some soft SUSY breaking deformation preserves certain abelian global symmetries of $\mathcal{N}=2$ SCFT, and the known anomaly of this preserved symmetry is quite useful in determining the IR phase:
a): For $\mathcal{N}=1$ preserving deformation, one can use it to determine the IR central charges and operator spectrum;  b): The deformation using bottom component of $\mathcal{N}=2$ supercurrent multiplet preserves $SU(2)_R\times U(1)_R$ symmetry, the IR theory has to be gapless to match the anomaly, which indicates that
the IR theory could be a non-supersymmetric CFT. 

This paper is organized as follows: section 2 classifies soft SUSY breaking deformation of $\mathcal{N}=1$ SCFT; section 3 classifies soft SUSY breaking deformation of $\mathcal{N}=2$ SCFT; section 
4 gave a more detailed study of $\mathcal{N}=1$ preserving deformation of $\mathcal{N}=2$ SCFTs; finally, a conclusion is given in section 5.

\section{Soft SUSY breaking of $\mathcal{N}=1$ SCFT}
Let's first consider soft SUSY breaking of four dimensional $\mathcal{N}=1$ SCFT. We begin with a short review of representation theory of $\mathcal{N}=1$ superconformal algebra.
The bosonic symmetry group of a $\mathcal{N}=1$ SCFT is $SO(2,4)\times U(1)_R \times G_F$, here $SO(2,4)$ 
is  conformal group of four dimensional Minkowski space time, $U(1)_R$ is the $R$ symmetry group which exists for every $\mathcal{N}=1$ SCFT, and $G_F$ are other continuous
global symmetry groups. A highest weight representation is labeled as $|\Delta,r, j_1, j_2\rangle$, where $\Delta$ is
the scaling dimension, $r$ is $U(1)_R$ charge, $j_1$ and $j_2$ are left and right spin. These states 
might also carry quantum numbers of flavor symmetry group $G_F$. 
Representation theory of $\mathcal{N}=1$ SCFT has been well studied and the short representation is classified in \cite{Flato:1983te,Dobrev:1985qv}. Two important short representations  are chiral multiplets 
and multiplets for conserved currents \footnote{The quantum numbers such as the scaling dimensions are the ones for the bottom component of the multiplet. }:
\begin{equation}
\begin{split}
& {\cal B}_{r,(j_1,0)},~~\Delta={3\over 2} r,~~~ \nonumber\\
& \hat{{\cal C}}_{(j_1,j_2)},~~~r=j_1-j_2,~~~\Delta=2+j_1+j_2.
\end{split}
\end{equation}
$\hat{{\cal C}}_{(0,0)}$ contains conserved currents for the global symmetry group $G_F$; $\hat{{\cal C}}_{({1\over2},0)}$ contains 
other supersymmetry currents; $\hat{{\cal C}}_{({1\over 2},{1\over2})}$ contains energy-moment tensor and $U(1)_R$ current, so
$\hat{{\cal C}}_{({1\over 2},{1\over2})}$ exists for any $\mathcal{N}=1$ SCFT. 
The anomaly of $U(1)_R$ symmetry is related to central charge $a$ and $c$ as follows \cite{Anselmi:1997am,Anselmi:1997ys} :
\begin{equation}
a={3\over 32}(3 \text{tr} R^3-\text{tr} R),~~c={1\over 32}(9 \text{tr} R^3-5 \text{tr} R).
\label{anom}
\end{equation}

Let's now consider deformation of $\mathcal{N}=1$ SCFT. We have following $F$ term relevant or marginal deformation \cite{Green:2010da}:
\begin{equation}
\int d^2\theta {\cal Z} {\cal B}_{r,(0,0)} +c.c,~~~r\leq 2.
\end{equation}
We can assign ${\cal Z}$ a scaling dimension $3-{3\over2}r$ so that above deformation is dimensionless. The soft condition is that ${\cal Z}$ has non-negative scaling dimension, which puts the constraint $r\leq 2$.
If ${\cal Z}$ is a constant, we have a \textbf{supersymmetric} relevant deformation for $r<2$, and exact marginal or marginal irrelevant depending on the global symmetry charge of operator \cite{Green:2010da}. If ${\cal Z}$ is promoted to 
a chiral superfield and all of its components are nonzero, then we can have soft supersymmetry breaking deformation. In particular, if we would 
just turn on the top component (with highest scaling dimension in the supermultiplet) of ${\cal Z}$, then we have 
most relevant supersymmetry breaking deformation from this chiral multiplet:
\begin{equation}
\boxed{\delta S=\lambda \int d^4 x  {\cal B}+c.c}.
\end{equation}
Here ${\cal B}$ is the bottom component of the chiral multiplet ${\cal B}_{r,(0,0)}$. 
This deformation breaks $U(1)_R$ symmetry, but it might preserve a combination of $U(1)_R$ symmetry and other anomaly free global symmetries. 
The anomaly of the preserved symmetry can be used to constrain the IR behavior. Similarly, one can use other chiral multiplet ${\cal B}_{r,(j_1,0)}$ with $j_1\neq 0$ to 
get soft SUSY breaking deformation, and the bound on the $r$ charge is just $r\leq 2$. 

If a $\mathcal{N}=1$ SCFT has other global symmetry group $G_F$,  we could  have following $D$ term deformation:
\begin{equation}
\delta S=\int d^2\theta d^2\bar{\theta} \Lambda \hat{{\cal C}}_{(0,0)}.
\end{equation}
We also assign scaling dimension zero to $\Lambda$ to make above deformation dimensionless. This is the only relevant or marginal deformation which can be derived from using $\hat{{\cal C}}$ type operator.  
Now if we promote $\Lambda$ to be a real superfield and assume only the top component of $\Lambda$ is nonzero, we have the following relevant deformation
\begin{equation}
\boxed{\delta S= m^2\int d^4x {\cal C}}.
\end{equation}
Here ${\cal C}$ is the bottom component of   $\hat{{\cal C}}_{(0,0)}$ and does not break $U(1)_R$ symmetry.
So if $U(1)_R$ symmetry is not spontaneously broken, the IR theory should match its anomaly, and we can potentially get
an interacting non-supersymmetric CFT in the IR by using above type of deformation.

\section{Soft SUSY breaking of $\mathcal{N}=2$ SCFT}
Let's now discuss soft SUSY breaking of four dimensional $\mathcal{N}=2$ SCFT, and we first review some representation theory results of $\mathcal{N}=2$ superconformal algebra. 
The bosonic symmetry group of a general $\mathcal{N}=2$ SCFT is $SO(2,4)\times SU(2)_R \times U(1)_R \times G_F$, here $SO(2,4)$ 
is the conformal group, $SU(2)_R\times U(1)_R$ is the $R$ symmetry group which exists for every $\mathcal{N}=2$ SCFT, and $G_F$ are other 
global symmetry groups which could be absent for some theories. A highest weight representation is labeled as $|\Delta,R,r, j_1, j_2\rangle$, here $\Delta$ is
the scaling dimension, $r$ is $U(1)_R$ charge, $R$ is $SU(2)_R$ charge, $j_1$ and $j_2$ are left and right spin. These states 
could also carry quantum numbers of flavor symmetry group $G_F$. Representation theory of $\mathcal{N}=2$ SCFT has been studied in \cite{Dolan:2002zh}.
and the short representation is completely classified in \cite{Dolan:2002zh}. Three short representations that we are interested in  are Coulomb branch operators, 
Higgs branch operators, and supercurrent multiplet:
\begin{equation}
\begin{split}
&\text{Coulomb~branch operators}:~~~~{\cal E}_{r,(0,0)},~~R=0,~~\Delta=r, \nonumber\\
&\text{Higgs~branch operators}:~~~~~~~~\hat{{\cal B}}_R,~~~~~~~r=j_1=j_2=0,~~\Delta=2R, \nonumber\\
&\text{Supercurrent}:~~~~~~~~~~~~~~~~~~~~~\hat{{\cal C}}_{0,(0,0)},~~r=R=0,~~\Delta=2. \nonumber\\
\end{split}
\end{equation}
$\hat{{\cal B}}_1$ is a multiplet which contains conserved current for the flavor symmetry group $G_F$, and  transforms in adjoint representation of $G_F$.
The $\mathcal{N}=1$ subalgebra is generated by the supercharge $Q_1$, and the corresponding $R$ symmetry is $R_{\mathcal{N}=1}={2\over 3} R_{\mathcal{N}=2}+{4\over 3} I_3$ \footnote{Here $R_{\mathcal{N}=2}$ is the generator for $\mathcal{N}=2$ $U(1)_R$ symmetry, and $I_3$ is the Cartan subalgebra of Lie algebra associated with $SU(2)_R$ symmetry. }. The other global 
symmetry group in $\mathcal{N}=1$  description is $J=2R_{N=2}-2I_3$ which commutes with the supercharge $Q_1$ \footnote{Our normalization is that $(Q_1, Q_2)$ 
are $SU(2)$ doublet with $I_3(Q_1)=-{1\over2}, I_3(Q_2)={1\over2}$, and $U(1)_R$ charges are  $R(Q_1)=R(Q_2)=-{1\over2}$.}.

Supercurrent multiplet exists for every $\mathcal{N}=2$ SCFT. 
There are also a large class of $\mathcal{N}=2$ SCFTs whose full Coulomb branch spectrum and Higgs branch spectrum are known \cite{Gaiotto:2009we,Xie:2012hs,Wang:2015mra,Wang:2018gvb,Xie:2015rpa}. 
We also know the central charge $a_{\mathcal{N}=2}$ and $c_{\mathcal{N}=2}$ which are related to the anomalies of $R$ symmetries as follows \cite{Shapere:2008zf}:
\begin{equation}
\begin{split}
&\text{Tr}(R_{\mathcal{N}=2}^3)=6(a_{\mathcal{N}=2}-c_{\mathcal{N}=2}),~~\text{Tr}(R_{\mathcal{N}=2})=24(a_{\mathcal{N}=2}-c_{\mathcal{N}=2}), \\
&\text{Tr}(R_{\mathcal{N}=2}R^2_{SU(2)})=(2a_{\mathcal{N}=2}-c_{\mathcal{N}=2}).
\end{split}
\end{equation}
So if a deformation preserves a subgroup $U(1)_{IR}=xR_{\mathcal{N}=2}+yI_3$ of $\mathcal{N}=2$ $U(1)_R\times SU(2)_R$ symmetry group, and the anomaly of $U(1)_{IR}$ can be computed as follows:
\begin{equation}
\begin{split}
&Tr(U(1)_{IR}^3)=6x^3(a_{\mathcal{N}=2}-c_{\mathcal{N}=2})+3xy^2(2a_{\mathcal{N}=2}-c_{\mathcal{N}=2}),\nonumber\\
&Tr(U(1)_{IR})=24x (a_{\mathcal{N}=2}-c_{\mathcal{N}=2}).
\end{split}
\label{iru1anom}
\end{equation}
If a deformation preserves $\mathcal{N}=1$ supersymmetry and a candidate $U(1)_R$ symmetry which is a linear combination of $\mathcal{N}=2$ R symmetry, one can use formula \ref{anom} and \ref{iru1anom} to compute 
the IR central charge:
\begin{equation}
\begin{split}
&a_{N=1}=(a_{N=2}-c_{N=2})[{27\over16}x^3-{9\over4}x]+(2a_{N=2}-c_{N=2})[{27\over32}xy^2], \nonumber\\
&c_{N=1}=(a_{N=2}-c_{N=2})[{27\over16}x^3-{15\over4}x]+(2a_{N=2}-c_{N=2})[{27\over32}xy^2].
\end{split}
\label{ircentral}
\end{equation}

\begin{table}
	\begin{center}
		\begin{tabular}{|c|c|c|c|c|c|c|}
			\hline
			~&$A$&$\Phi_i$& $B_{ij}$&$F^{\alpha \beta}$& $\Lambda_i$ & $C$\\ \hline
			$U(1)_R$ & $r$ & $r-{1\over2}$ & $r-1$&$r-1$&$r-{3\over 2}$ & $r-2$ \\ \hline
			$SU(2)_R$ & $0$ & ${1\over2}$  & $1$ & 0 & ${1\over2}$ & 0   \\ \hline	
			$\Delta$ & $r$ & $r+{1\over2}$ & $r+1$ & $r+1$&$r+{3\over 2}$ & $r+2$  \\ \hline		
		\end{tabular}
		\caption{Components of ${\cal E}_{r,(0,0)}$ multiplet. 
		Here $A$, $B_{ij}$ and  $C$ are scalars, $\Phi_i$ and $\Lambda_i$ are spinors. $F^{\alpha \beta}=\sigma_{ab}^{\alpha\beta}F^{ab-}$ with $F^{ab-}$ an antisymmetric anti-selfdual tensor. $B_{ij}$ is symmetric in $i,j$ index and transform in adjoint representation of $SU(2)_R$ group, and $\Phi_i$ and $\Lambda_i$ transforms in fundamental representation of $SU(2)_R$ group.}
		\label{chiral}
	\end{center}
\end{table}

\begin{table}
	\begin{center}
		\begin{tabular}{|c|c|c|c|c|}
			\hline
			~&$L^{\langle ij\rangle}$&$L_\alpha^{\langle i \rangle}$& $L_0$&$L_\mu$\\ \hline
			$U(1)_R$ & $0$ & $-{1\over2}$ & $-1$&$0$ \\ \hline
			$SU(2)_R$ & $1$ & ${1\over2}$  & $0$ & 0    \\ \hline	
			$\Delta$ & $2$ & ${5\over2}$ & $3$ & $3$  \\ \hline		
		\end{tabular}
		\caption{Components of a $\hat{{\cal B}}_1$ multiplet, here $L^{\langle ij\rangle}$ satisfies a reality condition. $L_\alpha^{\langle i \rangle}$ are a doublet of $SU(2)_R$ symmetry and is a spinor. $L_0$ is a complex scalar, and $L_\mu$ is a vector. }
		\label{higgs}
	\end{center}
\end{table}

The $\mathcal{N}=2$ preserving relevant or marginal deformations  have been completely classified in \cite{Argyres:2015ffa,Cordova:2016xhm}, and we have:
\begin{enumerate}
	\item Deformation using Coulomb branch operator:
	\begin{equation}
\delta S  = \int d^2 \theta_1d^2\theta_2 {\cal Z} {\cal E}_{r,(0,0)}+c.c,~~~~~~r \leq 2
	\end{equation}
 ${\cal E}_{r,(0,0)}$ type multiplet contains component fields $(A,\Phi_i, B_{ij}, F^{\alpha \beta}, \Lambda_i, C)$. Their quantum numbers under $R$ symmetry are 
listed in table. \ref{chiral}. Among five scalars, only $I_3(B_{11})=-1$ and $I_3(B_{22})=1$ carry non-trivial $SU(2)_R$ charge.  ${\cal Z}$ has scaling dimension $2-r$. 
If we promote ${\cal Z}$ to be a $\mathcal{N}=2$ chiral multiplet \footnote{$\mathcal{N}=2$ chiral multiplet has the same multiplet structure as the Coulomb branch operators ${\cal E}_{r,(0,0)}$.}, then we can have supersymmetry breaking deformation, see table. \ref{deform}
for deformations involving scalar operators in multiplet ${\cal E}_{r,(0,0)}$.

\begin{table}
	\begin{center}
		\begin{tabular}{|c||c|c|c|c|}
			\hline
			~& Deformation  &SUSY&Global symmetry& Scaling dimension\\ \hline
			${\cal E}_{r,(0,0)}$&$\delta S = \lambda  \int d^4x C +c.c$ & $\mathcal{N}=2$ & $SU(2)_R$ & $r+2$ \\ \hline
			~&$\delta S = \lambda \int  d^4x B_{11}+c.c$  &$\mathcal{N}=1$ 	& ${2\over r}U(1)_R+({2-{2\over r}})I_3$ & $r+1$ \\ \hline
			~&$\delta S = \lambda \int d^4x B_{22}+c.c$& $\mathcal{N}=1$ & ${2\over r}U(1)_R+({2\over r}-2)I_3$ & $r+1$ \\ \hline
			~&$\delta S = \lambda \int d^4x B_{12}+c.c$& $\mathcal{N}=0$ & $SU(2)_R$ & $r+1$ \\ \hline
			~&$\delta S = \lambda \int d^4x A$+c.c& $\mathcal{N}=0$ & $SU(2)_R$ & $r$ \\ \hline
			$\hat{{\cal B}}_1$&$\delta S  = m^2 \int d^4x L_0+c.c$ & $\mathcal{N}=2$ & $SU(2)_R$ & $3$ \\ \hline
			~&$\delta S  = m^2 \int d^4x L^{\langle 22 \rangle}+c.c$ & $\mathcal{N}=0$ & $ U(1)_R$ & $2$ \\ \hline
			~&$\delta S  = m^2 \int d^4x iL^{\langle 12 \rangle}$ & $\mathcal{N}=0$ & $SU(2)_R\times U(1)_R$ & $2$ \\ \hline
			$\hat{{\cal C}}_{0,(0,0)}$&$\delta S= m^2 \int d^4x J$ & $\mathcal{N}=0$ & $SU(2)_R\times U(1)_R$ & $2$ \\ \hline
		\end{tabular}
		\caption{$\mathcal{N}=2$ soft supersymmetry breaking deformations from  ${\cal E}_{r,(0,0)}$, $\hat{{\cal B}}_1$ and $\hat{{\cal C}}_{0,(0,0)}$ multiplets. We list the number of preserved SUSY, preserved global symmetry and the scaling dimension of the operator used in deformation.}
		\label{deform}
	\end{center}
\end{table}

\item Each $\hat{{\cal B}}_1$ operator contains fields $(L^{\langle ij\rangle}, L_\alpha^{\langle i \rangle}, L_0,L_\mu)$, see table. \ref{higgs} for their quantum numbers. There is a reality condition on fields $L^{\langle ij\rangle}$. This multiplet decomposes into a $\mathcal{N}=1$ 
chiral multiplet $X$ and a conserved current multiplet $L$, see table. \ref{higgs}. One can have a $\mathcal{N}=2$ preserving deformation:
	\begin{equation}
\delta S  = \int d^2 \theta_1 \Lambda X+c.c.
\end{equation}	
If we promote $\Lambda$ to be a superfield, we can have non-susy deformation.
\end{enumerate}

\begin{table}
\begin{center}
	\begin{tabular}{|c|c|c|c|}
		\hline
		~&${\cal O}$&$\lambda_{\alpha}$& $S$\\ \hline
		$U(1)_R$ & $r$ & $r-{1\over2}$ & $r-1$ \\ \hline
		$SU(2)_R$ & $0$ & $I_3(\lambda_\alpha)={1\over2}$  & $I_3(S)=1$  \\ \hline	
	\end{tabular}
\end{center}
\caption{The decomposition of a $\mathcal{N}=2$ Coulomb branch multiplet ${\cal E}_{r,(0,0)}$ into three $\mathcal{N}=1$ chiral multiplets: ${\cal O}$ and $S$ are of type ${\cal B}_{r,(0,0)}$, and $\lambda_{\alpha}$ is 
of type ${\cal B}_{r,({1\over2},0)}$.}
	\label{decom}
\end{table}

\begin{table}
\begin{center}
	\begin{tabular}{|c|c|c|}
		\hline
		~&$L$&$X$\\ \hline
		$U(1)_R$ & $0$ & $0$  \\ \hline
		$SU(2)_R$ & $0$ & $I_3(X)=1$    \\ \hline	
	\end{tabular}
	\label{Bonedecom}
\end{center}
\caption{The decomposition of a $\mathcal{N}=2$ $\hat{{\cal B}}_1$ multiplet into a $\mathcal{N}=1$ chiral multiplet $X$ and a $\mathcal{N}=1$ conserved current multiplet $L$.}
\label{flavor}
\end{table}

If we regard a $\mathcal{N}=2$ SCFT as a $\mathcal{N}=1$ SCFT (with supercharge $Q_1$), we have more choices for soft SUSY breaking. First, the constraint for operators with soft SUSY breaking is weakened, i.e. the bound on Coulomb branch operator $r$ charge 
is now $r\leq 3$. 
${\cal E}_{r,(0,0)}$ contains three $\mathcal{N}=1$ chiral multiplets, see table. \ref{decom}. The bottom component of $S$ is actually also
the top component of another $\mathcal{N}=1$ chiral if we choose a different supercharge $Q_2$. The top component of $S$ is also the top component of the 
whole $\mathcal{N}=2$ chiral multiplet,  so there is no new scalar SUSY breaking deformation from multiplet $S$.  Thus, the new interesting  SUSY breaking deformation comes from bottom component of $\mathcal{N}=1$ chiral multiplet ${\cal O}$ whose $r$ charge constraint is relaxed to be less than 3.
 $\hat{{\cal B}}_1$ multiplet contains a $\mathcal{N}=1$ chiral multiplet $X$ and a $\mathcal{N}=1$ conserved current multiplet $L$, and one can turn on SUSY breaking deformation using the current multiplet $L$. 

Moreover, the $\mathcal{N}=2$ current multiplet contains a 
$\mathcal{N}=1$ conserved current $\hat{J}$, a supersymmetry current $J_{\dot{\alpha}}$, and a supercurrent multiplet $J_{\alpha\dot{\alpha}}$.  
One can use the conserved current $\hat{J}$ multiplet to deform our theory: 
\begin{equation}
\delta S= m^2\int J,
\end{equation}
here $J$ is the bottom component of $\hat{{\cal C}}_{(0,0)}$ multiplet and transform trivially under $U(1)_R$ and $SU(2)_R$ symmetry, 
and $m^2$ has scaling dimension 2. 

In summary, for a $\mathcal{N}=2$ SCFT, we can have soft susy breaking deformation by using Coulomb branch multiplet ${\cal E}_{r,(0,0)}$ with $r\leq 2$, and a $\hat{{\cal B}}_1$ multiplet. 
If we regard our theory as a $\mathcal{N}=1$ SCFT, we can have soft susy breaking by using Coulomb branch multiplet ${\cal E}_{r,(0,0)}$ with $r\leq 3$,  $\hat{{\cal B}}_1$ multiplet, and 
$\hat{{\cal C}}_{0,(0,0)}$ multiplet, see table. \ref{deform}.

\textbf{Example}: Let's consider the deformations of the simplest Argyres-Douglas SCFT which is often called $(A_1, A_2)$ theory. This theory has following features:
\begin{itemize}
	\item The Coulomb branch spectrum is freely generated by an operator $u={\cal E}_{r,(0,0)}$ with $r={6\over 5}$. It does not have a Higgs branch, so there is no $\hat{{\cal B}}_1$ type multiplet.  There are only two Coulomb branch operators with $r\leq 3$: $u$ and $u^2$. 
	\item Its central charge is $a_{\mathcal{N}=2}={43\over 120}$, and $c_{\mathcal{N}=2}={11\over 30}$. 
\end{itemize}
The $\mathcal{N}=2$ soft SUSY breaking deformations are summarized in table. \ref{AD}. The $\mathcal{N}=1$ soft SUSY breaking deformations are summarized in table. \ref{AD1}.

The $\mathcal{N}=1$ preserving deformations were studied in \cite{Bolognesi:2015wta,Xie:2016hny,Buican:2016hnq}. We found three non-supersymmetric deformations with scaling dimension ${6\over 5}, {11\over 5}, {12\over 5}$ from using Coulomb branch operators, see table. \ref{AD} and \ref{AD1}. These deformations break the $U(1)_R$ symmetry and preserve $SU(2)_R$ symmetry, 
so we can not use anomaly matching to constrain its IR behavior. 
The  deformation using bottom component of $\hat{{\cal C}}_{0,(0,0)}$ 
preserves $U(1)_R\times SU(2)_R$ symmetry so the IR theory is constrained by the anomaly matching, and it should be a gapless theory in the IR. The IR theory is most likely an interacting CFT. 
The corresponding operator used in the deformation
has scaling dimension 2 which is smaller that the $\mathcal{N}=1$ deformation ([${\cal O}]={11\over 5}$) whose IR fixed point consists of $\mathcal{N}=1$ chiral scalar. So if the IR theory  after deformation using operator $J$ is indeed an interacting CFT, its central charge should be smaller than 
that of a  complex scalar and a Weyl fermion, which seems to be much smaller than the  known four dimensional non-supersymmetric CFT.  

\begin{table}
	\begin{center}
		\begin{tabular}{|c|c|c|c|c|c|}
			\hline
			Deformation &SUSY&Global symmetry& scaling dimension & $a$ & $c$\\ \hline
			$\delta S = \lambda  \int d^4x C^u+c.c $ & $\mathcal{N}=2$ & $SU(2)_R$ & ${16\over 5}$ & ${1\over 24}$ & ${1\over 6}$ \\ \hline
			$\delta S = \lambda \int  d^4x B_{11}^u+c.c$  &$\mathcal{N}=1$ 	& ${5\over3}U(1)_R+{1\over3}I_3$ & ${11\over 5}$&${1\over48}$&${1\over24}$ \\ \hline
			$\delta S = \lambda \int d^4x B_{22}^u+c.c$& $\mathcal{N}=1$ & ${5\over3}U(1)_R-{1\over 3}I_3$ & ${11\over 5}$&${1\over48}$&${1\over24}$ \\ \hline
			$\delta S = \lambda \int d^4x B_{12}^u+c.c$& $\mathcal{N}=0$ & $SU(2)_R$ & ${11\over 5}$ &N/A&N/A\\ \hline
			$\delta S = \lambda \int d^4x A^u+c.c$& $\mathcal{N}=0$ & $SU(2)_R$ & ${6\over5}$&N/A&N/A \\ \hline
		\end{tabular}
		\caption{Deformations using Coulomb branch operator $u$ of $(A_1, A_2)$ AD theory, and this comes from $\mathcal{N}=2$ soft supersymmetry breaking deformation.
		 For generic coupling constant of $\mathcal{N}=2$ preserving deformation, the IR theory is just a free $U(1)$ vector multiplet whose central charge is listed. For $\mathcal{N}=1$ preserving deformation, the 
		IR theory is conjectured to be a free $\mathcal{N}=1$ chiral multiplet \cite{Bolognesi:2015wta}. }
		\label{AD}
	\end{center}
\end{table}

\begin{table}
	\begin{center}
		\begin{tabular}{|c|c|c|c|c|c|}
			\hline
			Deformation &SUSY&Global symmetry& scaling dimension & $a$ & $c$\\ \hline
			$\delta S = \lambda \int  d^4x B_{11}^{u^2}+c.c$  &$\mathcal{N}=1$ 	& ${5\over6}U(1)_R+{7\over6}I_3$ & ${17\over 5}$&${263\over 768}$&${271\over 768}$\\ \hline
			$\delta S = \lambda \int d^4x A^{u^2}+c.c$& $\mathcal{N}=0$ & $SU(2)_R$ & ${12\over5}$&N/A&N/A\\ \hline
						$\delta S = m^2\int d^4x J$ & $\mathcal{N}=0$ & $SU(2)_R\times U(1)_R$ & $2$ & N/A & N/A \\ \hline
		\end{tabular}
		\caption{The first two rows summarizes the properties of deformations using  $\mathcal{N}=1$ chiral multiplet of Coulomb branch operator $u^2$. The third row summarizes the deformation using bottom component of $\mathcal{N}=2$ supercurrent multiplet. }
		\label{AD1}
	\end{center}
\end{table}

\newpage
\section{$\mathcal{N}=1$ preserving deformation of $\mathcal{N}=2$ SCFT}
Let's now consider more details of $\mathcal{N}=1$ preserving deformation of $\mathcal{N}=2$ SCFT.  We can consider deformations caused by Coulomb branch operators or Higgs branch operators.
The $\mathcal{N}=1$ deformations using $X$ component of $\hat{\cal B}_1$ operator preserves $SU(2)_R$ symmetry, and the IR theory actually has $\mathcal{N}=2$ SUSY whose 
behavior can be solved from Seiberg-Witten geometry. So we will mainly focus on deformation using Coulomb branch operators.  A  $\mathcal{N}=2$ Coulomb branch operator consists of 
three $\mathcal{N}=1$ chiral multiplets $({\cal O},\lambda_{\alpha}, S)$, and only operator ${\cal O}$ will give new $\mathcal{N}=1$ preserving deformations.

The Coulomb branch chiral ring is freely generated, and the generators can often be found from the Seiberg-Witten geometry.
One can turn on $\mathcal{N}=1$ preserving relevant deformations using operators from Coulomb branch chiral ring, see row two of table. \ref{deform}.  
Assuming we use a $\mathcal{N}=2$ Coulomb branch operator ${\cal E}_{r,(0,0)}$
with $R_{\mathcal{N}=2}$ charge $r$ to deform our theory:
\begin{equation}
\delta S =\lambda \int d^2\theta {\cal O}+c.c
\end{equation} 
Here ${\cal O}$ is the bottom $\mathcal{N}=1$ chiral multiplet of ${\cal E}_{r,(0,0)}$.
Then the candidate $U(1)_R$ symmetry for the IR theory is 
\begin{equation}
{2\over r}U(1)_R+({2-{2\over r}})I_3.
\label{iru1}
\end{equation}
The condition of relevant deformation on $r$ is simply $r<3$ \footnote{$r=3$ is marginal irrelevant, and the IR $U(1)_R$ symmetry is just the $U(1)_R$ symmetry of the UV $\mathcal{N}=2$ SCFT. Since the deformation  also carries the charge for the other global symmetry $J$, one can use the argument of \cite{Green:2010da} to conclude 
that this deformation is marginal irrelevant.}. 
A necessary condition for the above symmetry to be the true IR $U(1)_R$ symmetry is that all the chiral operators 
obey unitarity bound, which implies that all three $\mathcal{N}=1$ chiral multiplets contained in a $\mathcal{N}=2$ chiral  multiplet with minimal $R_{\mathcal{N}=2}$ charge $r_{min}$ 
should obey unitarity bound, and we have (see the quantum number of three $\mathcal{N}=1$ chiral multiplets in table. \ref{decom}.)
\begin{equation}
\begin{split}
& [{\cal O}_{min}]>1\rightarrow~{2r_{min}\over r}\times {3\over 2}>1, \nonumber \\
&[(\lambda_{\alpha})_{min}]>1\rightarrow~[{2(r_{min}-{1\over2})\over r}+(2-{2\over r}){1\over2}]\times {3\over2}>1, \nonumber\\
&[S_{min}]>1\rightarrow~ [{2(r_{min}-1)\over r}+(2-{2\over r})1]\times {3\over2}>1,
\end{split}
\end{equation}
and we find the following constraint:
 \begin{equation}
r>3-{3\over2}r_{min}.
\label{bound}
\end{equation}
If our theory contains Higgs branch operators $\hat{{\cal B}}_{1}$, then the scaling dimension of its $\mathcal{N}=1$ chiral $X$ would be 
\begin{equation}
[X]>1\rightarrow  (2-{2\over r})\times {3\over2}>1,
\end{equation} 
and we get the bound 
\begin{equation}
r>{3\over2}.
\label{bound1}
\end{equation}
This bound is larger than that from using Coulomb branch operator, so we use this bound if our theory has a  $\hat{{\cal B}}_{1}$ type operator. In general, if our theory has abelian flavor symmetries, the true IR $U(1)_R$ symmetry 
might also has a mixing with them. The Coulomb branch operators, however, is not charged under those global symmetries, and therefore 
would not detect the mixings. 

 For the Lagrangian theory or class ${\cal S}$ theory engineered using 
only regular punctures, the Coulomb branch operators have integral scaling dimension, and the only Coulomb branch operator that we can use to deform our theory 
is the one with scaling dimension two. Dimension two Coulomb branch operators give us $\mathcal{N}=2$ exact marginal deformations, and one can often write down 
a weakly coupled gauge theory descriptions where the dimensional two operators are constructed from $\mathcal{N}=2$ vector multiplet. The above $\mathcal{N}=1$ preserving deformation 
is just the mass deformation for $\mathcal{N}=1$ chiral multiplet inside a $\mathcal{N}=2$ vector multiplet. The situation becomes a lot more interesting for more general 
Argyres-Douglas theories where the Coulomb branch spectrum contains operators whose scaling dimension can be arbitrary distributed above the unitarity bound which is one for four dimensional scalar. 

Some properties of the IR theory can be found as follows (Assuming that the candidate IR $U(1)_R$ symmetry \ref{iru1} is the true $U(1)_R$ symmetry of the IR SCFT, and we consider only the Coulomb type deformations):
\begin{enumerate}
\item \textbf{Central charges}: One can compute the IR central charge using formula \ref{ircentral}, and we have $x={2\over r},~y=2-{2\over r}$.
\item \textbf{Index}: The explicit form of $\mathcal{N}=2$ Schur index of many interesting theories is known \cite{Xie:2019zlb}, and this index is actually invariant under RG flow, and can be used to get some useful information of IR theory \cite{Buican:2016hnq}.
\item \textbf{Chiral ring}: The $\mathcal{N}=1$ chiral operator ${\cal O}$ which is used to deform our theory satisfying a chiral ring relation ${\cal O}=0$ in the IR theory \cite{Xie:2016hny,Buican:2016hnq}. 
\item \textbf{Chiral spectrum}: One can get some information of chiral operators  from underlying $\mathcal{N}=2$ theory.  For a $\mathcal{N}=2$ chiral with $R_{N=2}$ charge $a$, the scaling dimension of its three $\mathcal{N}=1$ chiral multiplets are
\begin{equation}
[{\cal O}]={3a\over r},~~[\lambda_{\alpha}]={{3\over2}r-3+3a\over r},~~~[S]={3(r+a-2)\over r}.
\end{equation}
 If $2<r<3$, the minimal scaling dimension of a chiral scalar operator is $\Delta_{min}= {3r_{min}\over r}$;
if $1<r\leq 2$, we have $\Delta_{min}= {3(r+r_{min}-2)\over r}$. 
In particular, some of the chiral operators are relevant, and one 
can use them to deform IR $\mathcal{N}=1$ fixed point and flow to possibly new  $\mathcal{N}=1$ SCFT, although these deformations would break R symmetry, and we have little to say about IR theory.  The flavor symmetry  of UV $\mathcal{N}=2$ theory
is not broken by the Coulomb branch type $\mathcal{N}=1$ preserving deformations, so we do know the existence of $\hat{\cal C}_{(0,0)}$ type operators of IR $\mathcal{N}=1$ SCFT from the flavor symmetry of UV theory.

\item \textbf{Exact marginal deformations}: 1): If the UV $\mathcal{N}=2$ theory has a multiple number of Coulomb branch operators with $R_{\mathcal{N}=2}$ charge $r$, 
then the IR theory would have exact marginal deformations; 
2): if  UV $\mathcal{N}=2$ theory has a $\mathcal{N}=2$ exact marginal operator, then the $S$ component of it would be exact marginal in the IR $\mathcal{N}=1$ SCFT. 
3): If $r=2$, then the $\hat{{\cal B}}_2$ type operator might also give exact marginal deformations. 

\item \textbf{Inherited $\mathcal{N}=1$ duality}: If 4d $\mathcal{N}=2$ has an exact marginal deformation and has different duality frames,  then the IR theory would also have different duality frames. There are many $\mathcal{N}=2$ theories whose duality frames are known \cite{Xie:2016uqq,Xie:2017vaf,Xie:2017aqx}, and using the $\mathcal{N}=1$ preserving deformations, we 
get a large class of new type of Seiberg duality for $\mathcal{N}=1$ SCFTs.
\end{enumerate}

\begin{figure}
\begin{center}
\tikzset{every picture/.style={line width=0.75pt}} 

\begin{tikzpicture}[x=0.75pt,y=0.75pt,yscale=-1,xscale=1]

\draw    (341.07,170.07) -- (368.63,202.62) ;
\draw [shift={(368.63,202.62)}, rotate = 49.75] [color={rgb, 255:red, 0; green, 0; blue, 0 }  ][fill={rgb, 255:red, 0; green, 0; blue, 0 }  ][line width=0.75]      (0, 0) circle [x radius= 3.35, y radius= 3.35]   ;

\draw    (308,131) -- (339.78,168.54) ;
\draw [shift={(341.07,170.07)}, rotate = 229.75] [color={rgb, 255:red, 0; green, 0; blue, 0 }  ][line width=0.75]    (10.93,-3.29) .. controls (6.95,-1.4) and (3.31,-0.3) .. (0,0) .. controls (3.31,0.3) and (6.95,1.4) .. (10.93,3.29)   ;
\draw [shift={(308,131)}, rotate = 49.75] [color={rgb, 255:red, 0; green, 0; blue, 0 }  ][fill={rgb, 255:red, 0; green, 0; blue, 0 }  ][line width=0.75]      (0, 0) circle [x radius= 3.35, y radius= 3.35]   ;

\draw    (388.72,167.68) -- (369.51,202.45) ;
\draw [shift={(369.51,202.45)}, rotate = 118.92] [color={rgb, 255:red, 0; green, 0; blue, 0 }  ][fill={rgb, 255:red, 0; green, 0; blue, 0 }  ][line width=0.75]      (0, 0) circle [x radius= 3.35, y radius= 3.35]   ;

\draw    (418.29,130.41) -- (389.96,166.11) ;
\draw [shift={(388.72,167.68)}, rotate = 308.43] [color={rgb, 255:red, 0; green, 0; blue, 0 }  ][line width=0.75]    (10.93,-3.29) .. controls (6.95,-1.4) and (3.31,-0.3) .. (0,0) .. controls (3.31,0.3) and (6.95,1.4) .. (10.93,3.29)   ;
\draw [shift={(418.29,130.41)}, rotate = 128.43] [color={rgb, 255:red, 0; green, 0; blue, 0 }  ][fill={rgb, 255:red, 0; green, 0; blue, 0 }  ][line width=0.75]      (0, 0) circle [x radius= 3.35, y radius= 3.35]   ;

\draw (315,175.5) node [scale=0.9]  {$\lambda \int O$};
\draw (426,175.5) node   {$\lambda \int O$};
\draw (380,223) node   {$\mathcal{N} =1$};
\draw (365,121) node  [align=left] {S dual};
\draw (286,123) node   {$T_{1}$};
\draw (448,123) node   {$T_{2}$};

\end{tikzpicture}

\end{center}
\caption{The parent $\mathcal{N}=2$  SCFT has an exact marginal deformation and so there are two different duality frames $T_1$ and $T_2$ where one can write down weakly coupled gauge theory descriptions. We turn on $\mathcal{N}=1$ preserving deformations which might 
have different descriptions in $T_1$ and $T_2$ frames, and they actually flow to the same IR theory. The IR theory also has an exact marginal deformation, which is inherited from parent $\mathcal{N}=2$ SCFT.  }
\label{sduality}
\end{figure}
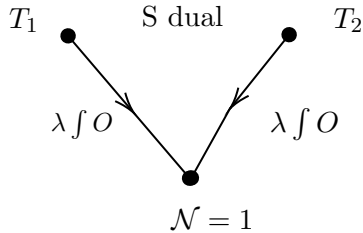

\textbf{Example}:  Let's consider $(A_1, G)$ theory with $G=ADE$, and these theories can be engineered by the following three-fold singularity \cite{Cecotti:2010fi}:
\begin{equation}
f(x,y,z,w)=f_{ADE}(x,y,z)+w^2.
\end{equation}
Here $f_{ADE}$ are standard two dimensional $ADE$ singularity with following form
\begin{equation}
\begin{split}
&A_{N}:~f=x^2+y^2+z^{N+1},~~~
D_{N}:~f=x^2+y^{N-1}+yz^2,~~~~\\
&E_6:~f=x^2+x^3+y^4,~~~ 
E_7:~f=x^2+x^3+xy^3,~~
E_8:~f=x^2+x^3+y^5.~~~
\end{split}
\end{equation}
The Coulomb branch spectrum of these theories can be computed using  the Jacobi algebra of the singularity \cite{Xie:2015rpa}. Let's review it here: the Jacobi algebra of an isolated singularity $f$ is defined as the following quotient space
\begin{equation}
J_f={C[x,y,z,w]\over \{ {\partial f\over \partial x},{\partial f\over \partial y}, {\partial f\over \partial z},{\partial f\over \partial w}\}}.
\end{equation}
Take a monomial basis $\phi_\alpha$ of $J_f$, then the Seiberg-Witten geometry of $f$ is 
\begin{equation}
F(x,y,z,w)=f(x,y,z,w)+\sum \lambda_{\alpha} \phi_{\alpha}.
\end{equation}
The singularity $f$ has a $C^*$ action: $f(\lambda^{q_i} z_i)=\lambda f(z_i)$, and the scaling dimension of $\lambda_\alpha$ is 
\begin{equation}
[\lambda_{\alpha}]={1-Q_\alpha\over \sum q_i-1}.
\end{equation} 
here $Q_\alpha$ is the weight of $\phi_\alpha$ under the $C^*$ action. 
The Coulomb branch chiral ring is freely generated by $\lambda_{\alpha}$ with $[\lambda_{\alpha}]>1$.  The central charges $a_{\mathcal{N}=2}$ and $c_{\mathcal{N}=2}$ are computed using the method 
in \cite{Xie:2015rpa}.  We also list the value $r>$ which is the lower bound given by formula \ref{bound} or \ref{bound1} (UV theory has $\hat{{\cal B}}_1$ type operators), 
and  the maximal possible value $r_{max}$ \footnote{The lower and maximal bound might not be realized in these theories.} that one can use to do a $\mathcal{N}=1$ preserving deformation. Using these operators, one can get a large class of interesting RG flow between a $\mathcal{N}=2$ SCFT
and a $\mathcal{N}=1$ SCFT.

\begin{table}
\begin{center}
	\begin{tabular}{|c|c|c|c|c|c|}
		\hline
		${\cal T}$&$a$&$c$&$r_{min}$&$r_{max}$&$r>$\\ \hline
		$(A_1, A_{2N})$ & ${N(24N+19)\over 24(2N+3)}$ & ${N(6N+5)\over 6(2N+3)}$ &${2N+4\over 2N+3}$&${6N+8\over 2N+3}$& ${3N+3\over 2N+3}$ \\ \hline
		$(A_1,A_{2N-1})$ & ${12N^2-5N-5\over 24(N+1)}$ & ${3N^2-N-1\over 6 (N+1)}$ & ${2N+4\over 2N+2}$& ${6N+4\over 2N+2}$& ${3\over 2}$ \\ \hline
		$(A_1, D_{2N+1})$ & ${N(8N+3)\over 8(2N+1)}$ & ${N\over 2}$ &${2N+2\over 2N+1}$&${6N+2\over 2N+1}$&${3\over 2}$\\ \hline
		$(A_1, D_{2N})$ & ${N\over2}-{5\over12}$ & ${N\over 2}-{1\over3}$ &${N+1\over N}$&${3N-1\over N}$& ${3\over 2}$ \\ \hline
		$(A_1, E_{6})$ & ${75\over 56}$ & ${19\over 14}$ &${8\over 7}$&${20\over 7}$ & ${9\over 7}$\\ \hline
		$(A_1, E_{7})$ & ${3\over 2}$ & ${31\over 20}$ &${6\over 5}$&${18\over 5}$ & ${3\over 2}$\\ \hline
		$(A_1, E_{8})$ & ${91\over 48}$ & ${23\over 12}$ &${9\over 8}$&${22\over 8}$ &  ${21\over16}$\\ \hline
	\end{tabular}
	\label{ade}
\end{center}
\caption{The central charges for $(A_1,G)$ type Argyres-Douglas theories. We listed the minimal scaling dimension $r_{min}$ of Coulomb branch operators, lower bound $r>$ and upper bound $r_{max}$ so that the corresponding $\mathcal{N}=1$ preserving deformations satisfies the unitarity constraint.  }
\label{a1g}
\end{table}

\textbf{Remark 1: Other $\mathcal{N}=1$ deformation and accidental symmetry}:  We have restricted our consideration to  $\mathcal{N}=1$ preserving deformation where no chiral operators violate unitarity bound under the candidate  IR $U(1)_R$ symmetry. We could also consider other deformations, and a common procedure is to 
assume these operators violating unitary bound to become free \cite{Kutasov:2003iy}. It would be interesting to further study these flows too. 

\textbf{Remark 2: New $\mathcal{N}=1$ duality}:  We have discussed $\mathcal{N}=1$ duality which is inherited from $\mathcal{N}=2$ duality. Here we show that it might be possible to find new $\mathcal{N}=1$ duality through following process: Let's start with a $\mathcal{N}=2$ SCFT 
and consider two $\mathcal{N}=1$ scalar chiral multiplets ${\cal O}_1$ and ${\cal O}_2$ (These are the bottom $\mathcal{N}=1$ chirals inside $\mathcal{N}=2$ Coulomb branch operators). The $R_{\mathcal{N}=2}$ charges of ${\cal O}_1$ and ${\cal O}_2$  are chosen to be different. Let's now consider following flow:
\begin{equation}
\delta S= \lambda_1\int d^2\theta {\cal O}_1+\lambda_2 \int d^2\theta {\cal O}_2+c.c.
\end{equation}
Let's denote our original $\mathcal{N}=2$ SCFT as theory $A$, and assume that the IR theory is a $\mathcal{N}=1$ SCFT with label $D$. The above flows might be considered in two steps, and there are two choices in first step: we can first use operator ${\cal O}_1$ to flow to theory $B$ and then use operator ${\cal O}_2$ to flow to theory $D$; or we can first use 
operator ${\cal O}_2$ to flow to a theory $C$, and then use operator ${\cal O}_1$ to flow to theory $D$. So we have established a $\mathcal{N}=1$ duality for theory $B$ and $C$, and this is quite similar to Seiberg duality \cite{Seiberg:1994pq}, and is manifest using the parent $\mathcal{N}=2$ description. See figure. \ref{flow}.

\textbf{Remark 3: Turning on expectation values}: Up to this point, we focus on the IR behavior of the origin of $\mathcal{N}=2$ moduli space ($\mathcal{N}=2$ SCFT point) after the relevant $\mathcal{N}=1$ preserving deformation, and it is interesting to consider the IR behavior of other points on the moduli space after the deformation. We leave the general study of this question to the future.

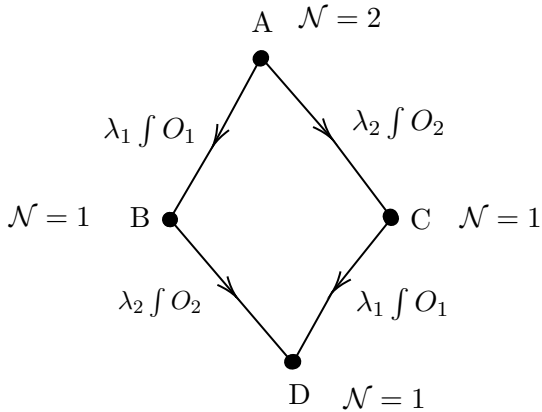
\begin{figure}[h]
\begin{center}
\tikzset{every picture/.style={line width=0.75pt}} 

\begin{tikzpicture}[x=0.75pt,y=0.75pt,yscale=-1,xscale=1]

\draw    (388.86,90.57) -- (417.79,129.33) ;
\draw [shift={(417.79,129.33)}, rotate = 53.26] [color={rgb, 255:red, 0; green, 0; blue, 0 }  ][fill={rgb, 255:red, 0; green, 0; blue, 0 }  ][line width=0.75]      (0, 0) circle [x radius= 3.35, y radius= 3.35]   ;

\draw    (353.49,49.57) -- (387.55,89.06) ;
\draw [shift={(388.86,90.57)}, rotate = 229.22] [color={rgb, 255:red, 0; green, 0; blue, 0 }  ][line width=0.75]    (10.93,-3.29) .. controls (6.95,-1.4) and (3.31,-0.3) .. (0,0) .. controls (3.31,0.3) and (6.95,1.4) .. (10.93,3.29)   ;
\draw [shift={(353.49,49.57)}, rotate = 49.22] [color={rgb, 255:red, 0; green, 0; blue, 0 }  ][fill={rgb, 255:red, 0; green, 0; blue, 0 }  ][line width=0.75]      (0, 0) circle [x radius= 3.35, y radius= 3.35]   ;

\draw    (329.22,94.09) -- (307.74,130.99) ;
\draw [shift={(307.74,130.99)}, rotate = 120.2] [color={rgb, 255:red, 0; green, 0; blue, 0 }  ][fill={rgb, 255:red, 0; green, 0; blue, 0 }  ][line width=0.75]      (0, 0) circle [x radius= 3.35, y radius= 3.35]   ;

\draw    (352.89,50.21) -- (330.17,92.33) ;
\draw [shift={(329.22,94.09)}, rotate = 298.34000000000003] [color={rgb, 255:red, 0; green, 0; blue, 0 }  ][line width=0.75]    (10.93,-3.29) .. controls (6.95,-1.4) and (3.31,-0.3) .. (0,0) .. controls (3.31,0.3) and (6.95,1.4) .. (10.93,3.29)   ;
\draw [shift={(352.89,50.21)}, rotate = 118.34] [color={rgb, 255:red, 0; green, 0; blue, 0 }  ][fill={rgb, 255:red, 0; green, 0; blue, 0 }  ][line width=0.75]      (0, 0) circle [x radius= 3.35, y radius= 3.35]   ;

\draw    (341.07,170.07) -- (368.63,202.62) ;
\draw [shift={(368.63,202.62)}, rotate = 49.75] [color={rgb, 255:red, 0; green, 0; blue, 0 }  ][fill={rgb, 255:red, 0; green, 0; blue, 0 }  ][line width=0.75]      (0, 0) circle [x radius= 3.35, y radius= 3.35]   ;

\draw    (308,131) -- (339.78,168.54) ;
\draw [shift={(341.07,170.07)}, rotate = 229.75] [color={rgb, 255:red, 0; green, 0; blue, 0 }  ][line width=0.75]    (10.93,-3.29) .. controls (6.95,-1.4) and (3.31,-0.3) .. (0,0) .. controls (3.31,0.3) and (6.95,1.4) .. (10.93,3.29)   ;
\draw [shift={(308,131)}, rotate = 49.75] [color={rgb, 255:red, 0; green, 0; blue, 0 }  ][fill={rgb, 255:red, 0; green, 0; blue, 0 }  ][line width=0.75]      (0, 0) circle [x radius= 3.35, y radius= 3.35]   ;

\draw    (388.72,167.68) -- (369.51,202.45) ;
\draw [shift={(369.51,202.45)}, rotate = 118.92] [color={rgb, 255:red, 0; green, 0; blue, 0 }  ][fill={rgb, 255:red, 0; green, 0; blue, 0 }  ][line width=0.75]      (0, 0) circle [x radius= 3.35, y radius= 3.35]   ;

\draw    (418.29,130.41) -- (389.96,166.11) ;
\draw [shift={(388.72,167.68)}, rotate = 308.43] [color={rgb, 255:red, 0; green, 0; blue, 0 }  ][line width=0.75]    (10.93,-3.29) .. controls (6.95,-1.4) and (3.31,-0.3) .. (0,0) .. controls (3.31,0.3) and (6.95,1.4) .. (10.93,3.29)   ;
\draw [shift={(418.29,130.41)}, rotate = 128.43] [color={rgb, 255:red, 0; green, 0; blue, 0 }  ][fill={rgb, 255:red, 0; green, 0; blue, 0 }  ][line width=0.75]      (0, 0) circle [x radius= 3.35, y radius= 3.35]   ;

\draw (422,82.5) node [scale=1]  {$\lambda _{2}\int O_{2}$};
\draw (354,31.57) node  [align=left] {A};
\draw (293,130) node  [align=left] {B};
\draw (433,131) node  [align=left] {C};
\draw (372.5,219) node  [align=left] {D};
\draw (303,172.5) node [scale=0.9]  {$\lambda _{2}\int O_{2}$};
\draw (424,173.5) node   {$\lambda _{1}\int O_{1}$};
\draw (298,86.5) node   {$\lambda _{1}\int O_{1}$};
\draw (393,29) node   {$\mathcal{N} =2$};
\draw (248,129) node   {$\mathcal{N} =1$};
\draw (473,130) node   {$\mathcal{N} =1$};
\draw (415,221) node   {$\mathcal{N} =1$};

\end{tikzpicture}

\end{center}
\caption{A $\mathcal{N}=2$ theory  $A$ is deformed by two $\mathcal{N}=1$ chiral multiplets ${\cal O}_1$ and ${\cal O}_2$ to get a theory $D$. There are two paths to interpret this flow: a) We first use operator ${\cal O}_1$ to get a $\mathcal{N}=1$ theory $B$ and then use operator ${\cal O}_2$ to get theory $D$; 
b) We first use operator ${\cal O}_2$ to get a $\mathcal{N}=1$ theory $C$ and then use operator ${\cal O}_2$ to get theory $D$. Theory $B$ and $C$ are very distinct $\mathcal{N}=1$ SCFT, so from $\mathcal{N}=1$ point of view, we find a Seiberg-duality: theory $B$ with a relevant deformation flows to the same theory 
as theory $C$ deformed by a different relevant deformation.}
\label{flow}
\end{figure}

\section{Discussion}
The space of $\mathcal{N}=2$ SCFT is increased dramatically in last few years, and lots of important properties about these theories such as 
the space of BPS operators, central charges, weakly coupled gauge theory duality frames, etc are known.  Most of previous studies focuses on 
the properties of $\mathcal{N}=2$ preserving deformations such as Seiberg-Witten geometry of Coulomb branch and Higgs branch chiral ring. Given the vast amount of 
knowledge of these $\mathcal{N}=2$ theories, it is time to study more about supersymmetry breaking deformations.  In this paper, 
We use the classification of $\mathcal{N}=2$ and $\mathcal{N}=1$ preserving relevant or marginal deformations 
to classify soft SUSY breaking of general $\mathcal{N}=2$ SCFT.  

Given the rich spectrum of Coulomb branch operators of a general $\mathcal{N}=2$ SCFT, it is now possible to construct a large class of 
new four dimensional $\mathcal{N}=1$ SCFT.  Many interesting properties about these theories such as central charges, chiral spectrum, Seiberg duality can be derived from the parent $\mathcal{N}=2$ 
SCFT.

We can also consider many interesting non-supersymmetric (non-SUSY) deformations, but we do not know too much about the phase structure of the IR theory. 
The hope is that since the deformations are soft, one can use the information of $\mathcal{N}=2$ SCFT like Seiberg-Witten solution to constrain
the behavior of  IR theory, and we would like to further study this question in the future. 
For the deformation that preserves some global symmetries whose anomalies are known and nontrivial, 
one can use it to constrain the IR theory, i.e. the IR theory should be gapless to match the anomaly. So simply from anomaly matching, we have found 
a large class of new non-SUSY CFT. Our non-SUSY CFT seems quite different from 
those found from non-abelian gauge theory \footnote{Among few known examples are famous Banks-Zaks fixed point.}, and it is definitely interesting to further study them.

\section*{Acknowledgements}
DX would like to thank Wenbin Yan for helpful discussions. 
 DX is supported by Yau Mathematical Science Center at Tsinghua University.


\appendix


\bibliographystyle{jhep}

\bibliography{ADhigher}

\providecommand{\href}[2]{#2}\begingroup\raggedright\begin{thebibliography}{10}

\bibitem{Cordova:2016xhm}
C.~Cordova, T.~T. Dumitrescu, and K.~Intriligator, {\it {Deformations of
  Superconformal Theories}},  {\em JHEP} {\bf 11} (2016) 135,
  [\href{http://xxx.lanl.gov/abs/1602.0121}{{\tt arXiv:1602.0121}}].

\bibitem{Cachazo:2002ry}
F.~Cachazo, M.~R. Douglas, N.~Seiberg, and E.~Witten, {\it {Chiral rings and
  anomalies in supersymmetric gauge theory}},  {\em JHEP} {\bf 12} (2002) 071,
  [\href{http://xxx.lanl.gov/abs/hep-th/0211170}{{\tt hep-th/0211170}}].

\bibitem{Gaiotto:2009we}
D.~Gaiotto, {\it {N=2 dualities}},  {\em JHEP} {\bf 1208} (2012) 034,
  [\href{http://xxx.lanl.gov/abs/0904.2715}{{\tt arXiv:0904.2715}}].

\bibitem{Xie:2012hs}
D.~Xie, {\it {General Argyres-Douglas Theory}},  {\em JHEP} {\bf 1301} (2013)
  100, [\href{http://xxx.lanl.gov/abs/1204.2270}{{\tt arXiv:1204.2270}}].

\bibitem{Wang:2015mra}
Y.~Wang and D.~Xie, {\it {Classification of Argyres-Douglas theories from M5
  branes}},  \href{http://xxx.lanl.gov/abs/1509.0084}{{\tt arXiv:1509.0084}}.

\bibitem{Wang:2018gvb}
Y.~Wang and D.~Xie, {\it {Codimension-two defects and Argyres-Douglas theories
  from outer-automorphism twist in 6d $(2,0)$ theories}},
  \href{http://xxx.lanl.gov/abs/1805.0883}{{\tt arXiv:1805.0883}}.

\bibitem{Xie:2015rpa}
D.~Xie and S.-T. Yau, {\it {4d N=2 SCFT and singularity theory Part I:
  Classification}},  \href{http://xxx.lanl.gov/abs/1510.0132}{{\tt
  arXiv:1510.0132}}.

\bibitem{Argyres:1995jj}
P.~C. Argyres and M.~R. Douglas, {\it {New phenomena in SU(3) supersymmetric
  gauge theory}},  {\em Nucl. Phys.} {\bf B448} (1995) 93--126,
  [\href{http://xxx.lanl.gov/abs/hep-th/9505062}{{\tt hep-th/9505062}}].

\bibitem{Argyres:1995xn}
P.~C. Argyres, M.~R. Plesser, N.~Seiberg, and E.~Witten, {\it {New N=2
  superconformal field theories in four-dimensions}},  {\em Nucl. Phys.} {\bf
  B461} (1996) 71--84, [\href{http://xxx.lanl.gov/abs/hep-th/9511154}{{\tt
  hep-th/9511154}}].

\bibitem{Beem:2013sza}
C.~Beem, M.~Lemos, P.~Liendo, W.~Peelaers, L.~Rastelli, and B.~C. van Rees,
  {\it {Infinite Chiral Symmetry in Four Dimensions}},  {\em Commun. Math.
  Phys.} {\bf 336} (2015), no.~3 1359--1433,
  [\href{http://xxx.lanl.gov/abs/1312.5344}{{\tt arXiv:1312.5344}}].

\bibitem{Xie:2019zlb}
D.~Xie and W.~Yan, {\it {Schur sector of Argyres-Douglas theory and
  $W$-algebra}},  \href{http://xxx.lanl.gov/abs/1904.0909}{{\tt
  arXiv:1904.0909}}.

\bibitem{Benini:2009mz}
F.~Benini, Y.~Tachikawa, and B.~Wecht, {\it {Sicilian gauge theories and N=1
  dualities}},  {\em JHEP} {\bf 01} (2010) 088,
  [\href{http://xxx.lanl.gov/abs/0909.1327}{{\tt arXiv:0909.1327}}].

\bibitem{Gadde:2013fma}
A.~Gadde, K.~Maruyoshi, Y.~Tachikawa, and W.~Yan, {\it {New N=1 Dualities}},
  {\em JHEP} {\bf 06} (2013) 056,
  [\href{http://xxx.lanl.gov/abs/1303.0836}{{\tt arXiv:1303.0836}}].

\bibitem{Maruyoshi:2013hja}
K.~Maruyoshi, Y.~Tachikawa, W.~Yan, and K.~Yonekura, {\it {N=1 dynamics with
  $T_N$ theory}},  {\em JHEP} {\bf 10} (2013) 010,
  [\href{http://xxx.lanl.gov/abs/1305.5250}{{\tt arXiv:1305.5250}}].

\bibitem{Maruyoshi:2016aim}
K.~Maruyoshi and J.~Song, {\it {$ \mathcal{N}=1 $ deformations and RG flows of
  $ \mathcal{N}=2 $ SCFTs}},  {\em JHEP} {\bf 02} (2017) 075,
  [\href{http://xxx.lanl.gov/abs/1607.0428}{{\tt arXiv:1607.0428}}].

\bibitem{Agarwal:2016pjo}
P.~Agarwal, K.~Maruyoshi, and J.~Song, {\it {$ \mathcal{N} $ =1 Deformations
  and RG flows of $ \mathcal{N} $ =2 SCFTs, part II: non-principal
  deformations}},  {\em JHEP} {\bf 12} (2016) 103,
  [\href{http://xxx.lanl.gov/abs/1610.0531}{{\tt arXiv:1610.0531}}]. [Addendum:
  JHEP04,113(2017)].

\bibitem{Bolognesi:2015wta}
S.~Bolognesi, S.~Giacomelli, and K.~Konishi, {\it {$ \mathcal{N}=2 $
  Argyres-Douglas theories, $ \mathcal{N}=1 $ SQCD and Seiberg duality}},  {\em
  JHEP} {\bf 08} (2015) 131, [\href{http://xxx.lanl.gov/abs/1505.0580}{{\tt
  arXiv:1505.0580}}].

\bibitem{Xie:2016hny}
D.~Xie and K.~Yonekura, {\it {Search for a Minimal N=1 Superconformal Field
  Theory in 4D}},  {\em Phys. Rev. Lett.} {\bf 117} (2016), no.~1 011604,
  [\href{http://xxx.lanl.gov/abs/1602.0481}{{\tt arXiv:1602.0481}}].

\bibitem{Buican:2016hnq}
M.~Buican and T.~Nishinaka, {\it {Small deformation of a simple $\mathcal N=2$
  superconformal theory}},  {\em Phys. Rev.} {\bf D94} (2016), no.~12 125002,
  [\href{http://xxx.lanl.gov/abs/1602.0554}{{\tt arXiv:1602.0554}}].

\bibitem{Luty:2005sn}
M.~A. Luty, {\it {2004 TASI lectures on supersymmetry breaking}},  in {\em
  {Physics in D >= 4. Proceedings, Theoretical Advanced Study Institute in
  elementary particle physics, TASI 2004, Boulder, USA, June 6-July 2, 2004}},
  pp.~495--582, 2005.
\newblock \href{http://xxx.lanl.gov/abs/hep-th/0509029}{{\tt hep-th/0509029}}.

\bibitem{Flato:1983te}
M.~Flato and C.~Fronsdal, {\it {Representations of Conformal Supersymmetry}},
  {\em Lett. Math. Phys.} {\bf 8} (1984) 159.

\bibitem{Dobrev:1985qv}
V.~K. Dobrev and V.~B. Petkova, {\it {All Positive Energy Unitary Irreducible
  Representations of Extended Conformal Supersymmetry}},  {\em Phys. Lett.}
  {\bf 162B} (1985) 127--132.

\bibitem{Anselmi:1997am}
D.~Anselmi, D.~Z. Freedman, M.~T. Grisaru, and A.~A. Johansen, {\it
  {Nonperturbative formulas for central functions of supersymmetric gauge
  theories}},  {\em Nucl. Phys.} {\bf B526} (1998) 543--571,
  [\href{http://xxx.lanl.gov/abs/hep-th/9708042}{{\tt hep-th/9708042}}].

\bibitem{Anselmi:1997ys}
D.~Anselmi, J.~Erlich, D.~Z. Freedman, and A.~A. Johansen, {\it {Positivity
  constraints on anomalies in supersymmetric gauge theories}},  {\em Phys.
  Rev.} {\bf D57} (1998) 7570--7588,
  [\href{http://xxx.lanl.gov/abs/hep-th/9711035}{{\tt hep-th/9711035}}].

\bibitem{Green:2010da}
D.~Green, Z.~Komargodski, N.~Seiberg, Y.~Tachikawa, and B.~Wecht, {\it {Exactly
  Marginal Deformations and Global Symmetries}},  {\em JHEP} {\bf 06} (2010)
  106, [\href{http://xxx.lanl.gov/abs/1005.3546}{{\tt arXiv:1005.3546}}].

\bibitem{Dolan:2002zh}
F.~Dolan and H.~Osborn, {\it {On short and semi-short representations for
  four-dimensional superconformal symmetry}},  {\em Annals Phys.} {\bf 307}
  (2003) 41--89, [\href{http://xxx.lanl.gov/abs/hep-th/0209056}{{\tt
  hep-th/0209056}}].

\bibitem{Shapere:2008zf}
A.~D. Shapere and Y.~Tachikawa, {\it {Central charges of N=2 superconformal
  field theories in four dimensions}},  {\em JHEP} {\bf 09} (2008) 109,
  [\href{http://xxx.lanl.gov/abs/0804.1957}{{\tt arXiv:0804.1957}}].

\bibitem{Argyres:2015ffa}
P.~Argyres, M.~Lotito, Y.~Lü, and M.~Martone, {\it {Geometric constraints on
  the space of $ \mathcal{N} $ = 2 SCFTs. Part I: physical constraints on
  relevant deformations}},  {\em JHEP} {\bf 02} (2018) 001,
  [\href{http://xxx.lanl.gov/abs/1505.0481}{{\tt arXiv:1505.0481}}].

\bibitem{Xie:2016uqq}
D.~Xie and S.-T. Yau, {\it {New N = 2 dualities}},
  \href{http://xxx.lanl.gov/abs/1602.0352}{{\tt arXiv:1602.0352}}.

\bibitem{Xie:2017vaf}
D.~Xie and S.-T. Yau, {\it {Argyres-Douglas matter and N=2 dualities}},
  \href{http://xxx.lanl.gov/abs/1701.0112}{{\tt arXiv:1701.0112}}.

\bibitem{Xie:2017aqx}
D.~Xie and K.~Ye, {\it {Argyres-Douglas matter and S-duality: Part II}},  {\em
  JHEP} {\bf 03} (2018) 186, [\href{http://xxx.lanl.gov/abs/1711.0668}{{\tt
  arXiv:1711.0668}}].

\bibitem{Cecotti:2010fi}
S.~Cecotti, A.~Neitzke, and C.~Vafa, {\it {R-Twisting and 4d/2d
  Correspondences}},  \href{http://xxx.lanl.gov/abs/1006.3435}{{\tt
  arXiv:1006.3435}}.

\bibitem{Kutasov:2003iy}
D.~Kutasov, A.~Parnachev, and D.~A. Sahakyan, {\it {Central charges and U(1)(R)
  symmetries in N=1 superYang-Mills}},  {\em JHEP} {\bf 11} (2003) 013,
  [\href{http://xxx.lanl.gov/abs/hep-th/0308071}{{\tt hep-th/0308071}}].

\bibitem{Seiberg:1994pq}
N.~Seiberg, {\it {Electric - magnetic duality in supersymmetric nonAbelian
  gauge theories}},  {\em Nucl. Phys.} {\bf B435} (1995) 129--146,
  [\href{http://xxx.lanl.gov/abs/hep-th/9411149}{{\tt hep-th/9411149}}].

\bibitem{Maruyoshi:2016tqk}
K.~Maruyoshi and J.~Song, {\it {Enhancement of Supersymmetry via
  Renormalization Group Flow and the Superconformal Index}},  {\em Phys. Rev.
  Lett.} {\bf 118} (2017), no.~15 151602,
  [\href{http://xxx.lanl.gov/abs/1606.0563}{{\tt arXiv:1606.0563}}].

\bibitem{Agarwal:2017roi}
P.~Agarwal, A.~Sciarappa, and J.~Song, {\it {$ \mathcal{N} $ =1 Lagrangians for
  generalized Argyres-Douglas theories}},  {\em JHEP} {\bf 10} (2017) 211,
  [\href{http://xxx.lanl.gov/abs/1707.0475}{{\tt arXiv:1707.0475}}].

\bibitem{Benvenuti:2017bpg}
S.~Benvenuti and S.~Giacomelli, {\it {Lagrangians for generalized
  Argyres-Douglas theories}},  {\em JHEP} {\bf 10} (2017) 106,
  [\href{http://xxx.lanl.gov/abs/1707.0511}{{\tt arXiv:1707.0511}}].

\bibitem{Giacomelli:2017ckh}
S.~Giacomelli, {\it {RG flows with supersymmetry enhancement and geometric
  engineering}},  {\em JHEP} {\bf 06} (2018) 156,
  [\href{http://xxx.lanl.gov/abs/1710.0646}{{\tt arXiv:1710.0646}}].

\bibitem{Giacomelli:2018ziv}
S.~Giacomelli, {\it {Infrared enhancement of supersymmetry in four
  dimensions}},  {\em JHEP} {\bf 10} (2018) 041,
  [\href{http://xxx.lanl.gov/abs/1808.0059}{{\tt arXiv:1808.0059}}].

\end{thebibliography}\endgroup

\end{document}